\documentclass[fleqn,10pt]{wlscirep}

\usepackage{graphicx,epsfig}
\usepackage{times}
\usepackage{graphics,dcolumn,bm,fleqn,epic,eepic}
\usepackage{amssymb,amsmath,multirow,rotate,color}
\usepackage[caption=false]{subfig}
\usepackage{color}
\usepackage{soul}
\usepackage{tikz}
\usetikzlibrary{patterns}
\usepackage{pgfplots}

\title{A Comparative Analysis of Community Detection Algorithms on Artificial Networks}

\author[1,*]{Zhao Yang}
\author[1]{Ren\'{e} Algesheimer}
\author[1]{Claudio J.~Tessone}
\affil[1]{URPP Social Networks, University of Z\"urich, Andreasstrasse 15, CH-8050 Z\"urich, Switzerland}

\affil[*]{zhao.yang@business.uzh.ch}

\begin{abstract}
Many community detection algorithms have been developed to uncover the mesoscopic properties of complex networks. However how good an algorithm is, in terms of accuracy and computing time, remains still open. Testing algorithms on real-world network has certain restrictions which made their insights potentially biased: the networks are usually small, and the underlying communities are not defined objectively. In this study, we employ the Lancichinetti-Fortunato-Radicchi benchmark graph to test eight state-of-the-art algorithms. We quantify the accuracy using complementary measures and algorithms' computing time. Based on simple network properties and the aforementioned results, we provide guidelines that help to choose the most adequate community detection algorithm for a given network. Moreover, these rules allow uncovering limitations in the use of specific algorithms given macroscopic network properties. 
Our contribution is threefold: firstly, we provide actual techniques to determine which is the most suited algorithm in most circumstances based on observable properties of the network under consideration. Secondly, we use the mixing parameter as an easily measurable indicator of finding the ranges of reliability of the different algorithms. Finally, we study the dependency with network size focusing on both the algorithm's predicting power and the effective computing time.

\end{abstract}

\begin{document}
\maketitle

\section*{Introduction}

Relationships between constituents of complex systems (be it in nature, society, or technological applications) can be represented in terms of networks. In this portrayal, the elements composing the system are described as nodes and their interactions as links. At the global level, the topology of these interactions -- far from being trivial -- is in itself of complex nature 
\cite{newman2003structure, boccaletti2006complex}. 
Importantly, these networks further display some level of organisation at an intermediate scale. 
At this \textit{mesoscopic} level, it is possible to identify groups of nodes that are heavily connected among themselves, but sparsely connected to the rest of the network. These interconnected groups are often characterised as \textit{communities}, or in other contexts \textit{modules}, and occur in a wide variety of networked systems \cite{girvan2002community, fortunato2010community}. 

Detecting communities has grown into a fundamental, and highly relevant problem in network science with multiple applications.
First, it allows to unveil the existence of a non-trivial internal network organisation at coarse grain level. 
This allows further to infer special relationships between the nodes that may not be easily accessible from direct empirical tests 
\cite{lancichinetti2008benchmark}. 
Second, it helps to better understand the properties of dynamic processes taking place in a network. 
As paradigmatic examples, spreading processes of epidemics and innovation are considerably affected by the community structure of the graph 
\cite{lancichinetti2010characterizing}.

Taking into account its importance, it is not surprising that many community detection methods have been developed, using tools and techniques from variegated disciplines such as statistical physics, biology, applied mathematics, computer science, and sociology. 
All these methods aim at improving the identification of meaningful communities, while keeping as low as possible the computational complexity of the underlying algorithm. 
Clearly, these algorithms are based on slightly different definitions of community, and therefore the results are not always directly comparable.
Further, in most real-world applications, a \textit{ground truth} -- i.e.~a \textit{unique} identification of nodes to communities -- is simply non-existent, which makes it even more difficult to assess the reliability of the community detection procedures. To address these shortcomings and test the algorithms' reliability, different benchmarks have been developed. 

Essentially, testing a community detection algorithm implies analysing computer-generated or real-world networks with a well defined community structure (a known ground truth) in order to obtain the community decomposition. 
One of the most used techniques is the GN benchmark (for Girvan \& Newman 
\cite{girvan2002community}), which is a special case of the planted $l-$partition model  
\cite{condon2001algorithms} with a prior specification of the number of nodes (128) and equally sized communities (4). 
When the expected number of links joining a node to others in different groups is smaller than 8, the four groups are strongly defined communities. 
In these conditions, a well functioning detection algorithm should be able to identify the communities in reasonable time. 
Different community detection algorithms can be compared based on their performances on the GN benchmark, which has already been done by Danon \textit{et al.} 
\cite{danon2005comparing}. 
However, there are several drawbacks to the GN benchmark: All nodes have the same expected degree, communities are separated in the same way, and the network is of an unrealistic small size. 

It is a well established fact that most real complex networks are characterised by largely heterogeneous degree distributions 
\cite{newman2003structure, boccaletti2006complex, barabasi1999emergence} and heterogeneous community sizes \cite{palla2005uncovering, guimera2003self, clauset2004finding}. 
For this reason, the GN benchmark cannot be considered as a good proxy for a real network. 
By consequence, in a newer stream of research 
\cite{lancichinetti2008benchmark, lancichinetti2009benchmarks}, the authors proposed an alternative benchmark, which is usually referred to as LFR (for Lancichinetti, Fortunato \& Radicchi). 
This method introduces power-law distributions of degree and community size to the graphs to generalise the GN benchmark. 
The performances of most existing community detection algorithms are good on the GN benchmark.
In contrast, the LFR benchmark presents a harder test for algorithms and makes it easier to unveil their limitations. It has been shown that the \textit{mixing parameter}, which is defined as
\begin{equation}
  \mu = \frac{\sum_i k^{ext}_i}{\sum_i k_i^{tot}}
  \label{mixingparameter}
\end{equation} 
is the most influential parameter in the LFR benchmark graphs
\cite{orman2009comparison}. 
Here $k^{ext}_i$ and $k^{tot}_i$ stand for the external degree of node $i$, i.e.~the number of edges connecting it to others that belong to different communities, and the total degree of said node.
Although it would be possible to define a mixing parameter for each node, it is assumed that $\mu$ is a global property and is the same for every node in the LFR benchmark. The reason here is to be consistent with the standard hypotheses of the planted $l$-partition model
\cite{lancichinetti2009community}.

According to the definition of community in a strong sense, each node should have more connections within the community than with the rest of the graph 
\cite{radicchi2004defining}.  Therefore, for $\mu > 1/2$ communities in the strong sense disappear. However, it is worth to mention that Lancichinetti and Fortunato \cite{lancichinetti2009community} found a weaker condition for community detection which can be applied to any version of the planted $l$-partition model:  
$\mu < (N-n^{max}_c)/N$, 
where $N$ is the total number of nodes, and $n^{max}_c$ is the size of the largest community. 
In our study, although we stick to the strong definition of communities, we have also taken the general condition of $\mu$ into consideration (see Table \ref{table1}).

In the following, we briefly review studies comparing community detection algorithms  in chronological order 
\cite{danon2005comparing, lancichinetti2008benchmark, lancichinetti2009benchmarks, lancichinetti2009community, orman2009comparison, peel2010estimating, hric2014community}  to highlight the research interests shift. In one of the early studies in comparing community detection algorithms, Danon \textit{et al.} had tested ten algorithms on the GN benchmark7
\cite{danon2005comparing} and collected estimates of how time complexity scales with network observables. 
However, the authors were not able to compare the actual computational effort as a result of the small sizes of graphs. Later on, Lancichinetti \textit{et al.} had employed the LFR benchmark to measure the accuracy of two algorithms on undirected unweighted networks without overlapping communities 
\cite{lancichinetti2008benchmark} and two algorithms on directed weighted networks with overlapping communities
\cite{lancichinetti2009benchmarks}. 
Concurrently, the authors tested twelve different algorithms on the GN and LFR benchmarks, and random graphs. 
For the tests on the LFR benchmark, the authors had considered various parameters, including undirected unweighted graphs with non-overlapping communities, directed unweighted graphs with non-overlapping communities, undirected weighted graphs with non-overlapping communities, and undirected unweighted graphs with overlapping communities 
\cite{lancichinetti2009community}.
Orman and Labatut later tested five community detection algorithms on the LFR benchmark
\cite{orman2009comparison}. They measured the accuracy of algorithms and studied the properties of the LFR benchmark graphs. 
Later, Peel applied two algorithms on both weighted and unweighted networks with 100 nodes and examined the performance of algorithms developed for weighted networks against those for unweighted ones for different parts of the problem space 
\cite{peel2010estimating}. Recently, Hric \textit{et al.} compared the accuracy of eleven different algorithms on both the LFR benchmark and a collection of real world graphs with sizes vary from 34 to 5189809 nodes
\cite{hric2014community}. Overall, as an extension of the GN benchmark, the LFR has drawn a lot of attention: Early, researchers employed small artificial and/or real world networks as benchmarks (e.g.~the GN benchmark and the Zachary's karate club network) ; while nowadays people shifted towards the use of large stylised large artificial or real world networks with some kind of ground truth obtained from metadata information (e.g.~the LFR benchmark and the DBLP collaboration network \cite{yang2015defining}). However, as of today, a detailed study of the dependency with the network size is missing as most of the existing studies include a few, selected, set of values of the number of nodes and the mixing parameter, and do not consider the real computing time needed to perform the analysis. 

In this paper, we evaluate eight different state-of-the-art community detection algorithms available in the ``igraph'' package \cite{igraphinfo}, which is a widely used collection of network analysis tools in R, Python, C and C++, on the LFR benchmark for undirected, unweighted graphs with non-overlapping communities. 
Details of the algorithms can be found in the methods section. 
Our contribution is threefold: First and foremost, we provide actual techniques to determine which is the most suited algorithm in most circumstances based on observable properties of the network under consideration. Secondly, we use the mixing parameter as an easily measurable indicator of finding the ranges of reliability of the different algorithms. Finally, we systematically study the dependency with network size focusing on both the algorithm's predicting power and the effective computing time.

\section*{Results}

In this section, we compare the results of community detection algorithms in terms of accuracy and computing time.
The former is defined as a measure of similarity between the modular structure generated by the LFR benchmark $\mathcal{P}$ (see Methods Section) and the partition identified by the respective community detection algorithms $\bar{\mathcal{P}}$. 
The latter is the real computing time needed to perform the community detection. 
This section is organised as follows: First, by employing the LFR generative model, we unveil the relationship between the mixing parameter and the accuracy of the community detection algorithms. 
Accuracy is measured in two different, complementary ways: The normalised mutual information \cite{danon2005comparing}, and the ratio between the number of detected communities and the number of communities given by the LFR generating model.
Then, we measure the computing time of community detection algorithms and show the relationship between the mixing parameter and the computing time. 
We then present the mixing parameter as computed from the communities detected by the different algorithms as a function of the input mixing parameter. 
Last, we present the comparisons of community detection algorithms in terms of accuracy and computing time as a function of network sizes.

\subsection*{The role of the network mixing parameter on accuracy and computing time}

First, we study the accuracy of the community detection algorithms as a function of the mixing parameter $\mu$. 
To measure the accuracy we have employed the \textit{normalised mutual information}, i.e., NMI. This is a measure borrowed from information theory which has been regularly used in papers comparing community detection algorithms \cite{lancichinetti2009benchmarks}.

Defining a \textit{confusion matrix} $\mathbf{N}$, where the rows correspond to the `real' communities, and the columns correspond to the `found' communities. The element of $\mathbf{N}$, $N_{ij}$, is the number of nodes in the real community $i$ that appear in the $j$-th detected community. The \textit{normalised mutual information} is then 
\cite{danon2005comparing}
 \begin{equation}
   I(\mathcal{P}, \bar{\mathcal{P}}) = \frac
   {-2\sum_{i=1}^{C}\sum_{j=1}^{\bar{C}}N_{ij}\log(N_{ij}N/N_{i \circ}N_{\circ j})}
   {\sum_{i=1}^{C} N_{i\circ}\log(N_{i\circ}/N)+\sum_{j=1}^{\bar{C}} N_{\circ j}\log(N_{\circ j}/N)}
    \label{nmi}
 \end{equation} 
where the number of communities given by the LFR model is denoted by $C$ and the number of communities detected by the algorithm is denoted by $\bar{C}$. The sum over the $i$-th row of $\mathbf{N}$ is denoted $N_{i\circ}$ and the sum over the $j$-th column is denoted $N_{\circ j}$. If the estimated communities are identical to the real ones, $I(\mathcal{P}, \bar{\mathcal{P}})$ equals to 1. If the partition found by the algorithm is totally independent from the real partition, $I(\mathcal{P}, \bar{\mathcal{P}})$ vanishes.

As pointed out in Ref.~ \cite{vinh2010information}, the mutual information can be normalised in different ways.  
These different normalisation methods are sensitive to different partition properties and have different theoretical properties 
\cite{vinh2010information, romano2014standardized, zhang2015evaluating}.
To get a better overview of the accuracy, we have calculated the NMI by using all these five different definitions (cf.~SI). We conclude that in the current study different normalisation procedures provide qualitatively similar behaviours. Just for the sake of brevity, and consistently with Danon \textit {et al.} 
\cite{danon2005comparing}, we report in this section only $I_{sum}$ (i.e.~normalisation by the arithmetic mean). The results of the other NMIs are shown in the ``Supplementary Information".

\begin{figure}
\caption{(lower row) The mean value of normalised mutual information depending on the mixing parameter $\mu$. (upper row) The standard deviation of the NMI as a function of $\mu$.}
\begin{center}
\includegraphics[scale = 0.08]{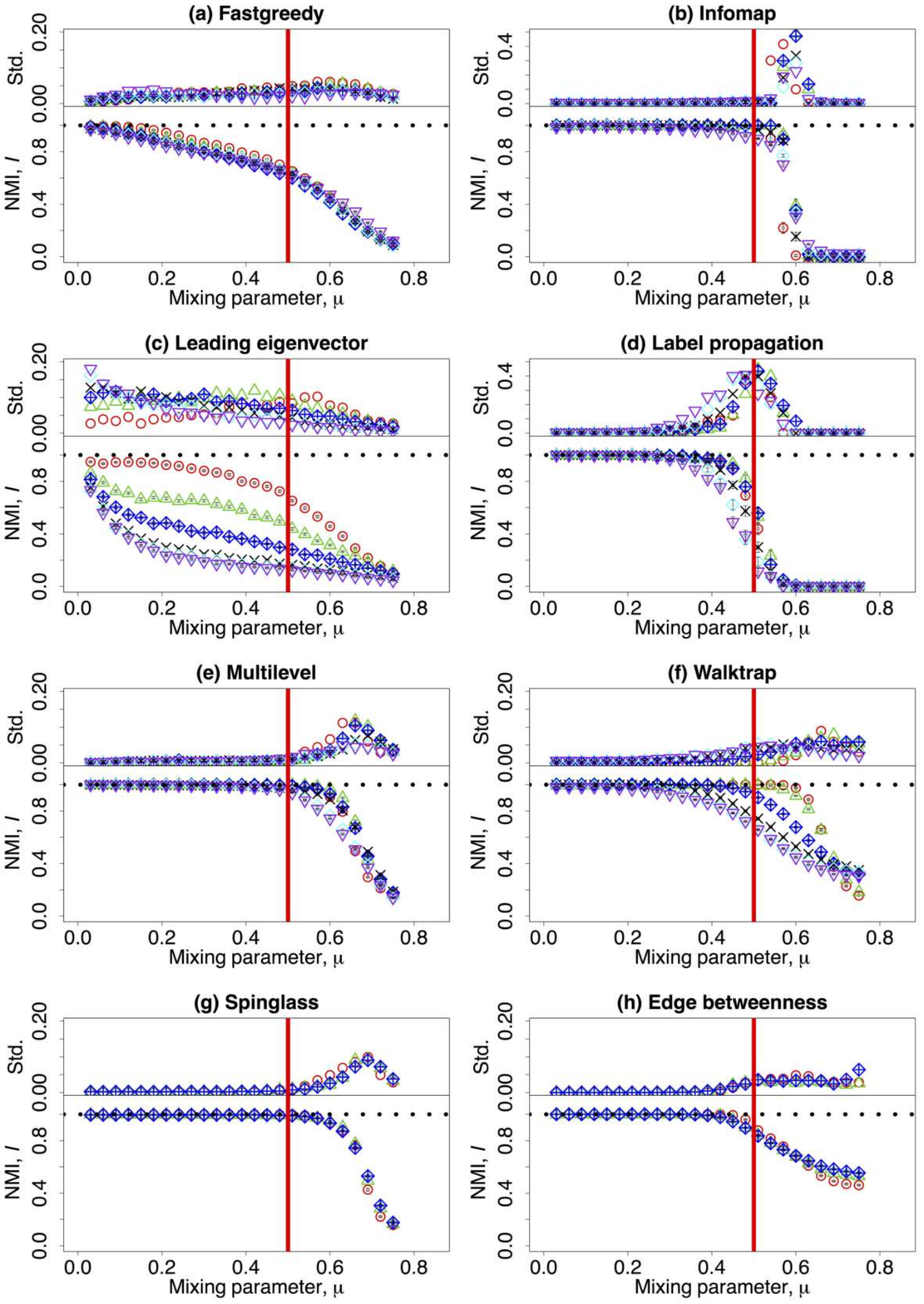}
\end{center}
\medskip
\small
Different colours refer to different number of nodes:  red ($N = 233$), green ($N = 482$), blue ($N = 1000$), black ($N = 3583$), cyan ($N = 8916$), and purple ($N = 22186$). Please notice that the vertical axis on the subfigures might have different scale ranges. The vertical red line corresponds to the strong definition of community, i.e.~$\mu = 0.5$. The horizontal black dotted line corresponds to the theoretical maximum, $I=1$. The other parameters are described in Table \ref{table1}.
\label{figure1}
\end{figure}

The results are shown in Figure \ref{figure1}. 
Each panel presents the accuracy of a given community detection algorithm and is subdivided into two plots: The lower axis depict the average value of NMI and the upper ones contain the standard deviation of the measures when repeated over 100 different network realisations. 
Most of the algorithms can uncover well the communities when the mixing parameter $\mu$ is small, as it is apparent from the large values of $I$ in the limit $\mu \to 0$. 
The accuracy of algorithms decreases, then, with increasing values of both network size and $\mu$. Different algorithms behave differently: the accuracy of Fastgreedy algorithm decreases monotonically, in a smooth fashion and has a very small standard deviation along all the range (Panel (a), Figure \ref{figure1}). Whereas that of Leading eigenvector algorithm falls rapidly even with small value of $\mu$ (Panel (c), Figure \ref{figure1}).
All the other algorithms display abrupt changes of behaviour: their performances remain relatively stable before a turning point where the NMI drops very fast as a function of $\mu$. 
The changes of behaviour are usually around $\mu = 1/2$, which corresponds to the strong definition of community
\cite{radicchi2004defining}. 
Interestingly, Label propagation and Edge betweenness algorithms have turning points smaller than said value; while Infomap, Multilevel, Walktrap, and Spinglass algorithms have turning points greater than $\mu = 1/2$. We have also noticed that for the Infomap algorithm the normalised mutual information has a point of discontinuous behaviour at around $\mu \cong 0.55$. On the other hand, for Label propagation, $I$ vanishes around $\mu \cong 0.5$ falling in a continuous fashion. This supports the conjecture that Infomap displays a first order phase transition as a function of the mixing parameter, while Label propagation algorithm may have a second order one. 
Nonetheless, we have not performed an exhaustive analysis on the matter to systematically analyse the existence (or not) of critical points. Further studies concerning the properties of these points are definitely needed.

Network size also plays the role here that a larger network size will lead to loss of accuracy at a lower value of $\mu$. 
For small enough networks ($N \leq 1000$), Infomap, Multilevel, Walktrap, and Spinglass outperform the other algorithms with higher values of $I$ and very small standard deviations, which shows the repeatability of the partitions detected. 
Besides, the turning point for accuracy is after $\mu = 1/2$. For larger networks ($N > 1000$), Infomap, Multilevel and Walktrap algorithms have relatively better accuracies and smaller standard deviations. Label propagation algorithm has much larger standard deviations such that its outputs are not stable. Due to the long computing time, Spinglass and Edge betweenness algorithms are too slow to be applied on large networks. 

Second, we study how well the community detection algorithms reproduce the number of communities. To do so, we compute the ratio $\bar{C}/C$ as a function of the mixing parameter. 
$\bar{C}$ is the average number of detected communities delivered by the different algorithms when repeated over 100 different network realisations. $C$ is the average real number of communities provided by the LFR benchmark on the same 100 networks. 
If $\bar{C} / C = 1$, the community detection algorithms are able to estimate correctly the number of communities. 
It is important to remark that this parameter has to be analysed together with the normalised mutual information because the distribution of community sizes is very heterogeneous.
With respect to the networks generated by the LFR model, for small network sizes the real number of communities is stable for all values of $\mu$, while for larger network sizes ($N > 1000$), $C$ grows up to $\mu \gtrapprox 0.2$ and then it saturates.

\begin{figure}
\caption{The mean value of the estimated number of communities delivered by different algorithms over the real number of communities given by the LFR benchmark, i.e., $\bar{C} / C$,  dependent on the mixing parameter $\mu$ on a \textit{log-linear} scale.}
\begin{center}
\includegraphics[scale = 0.08]{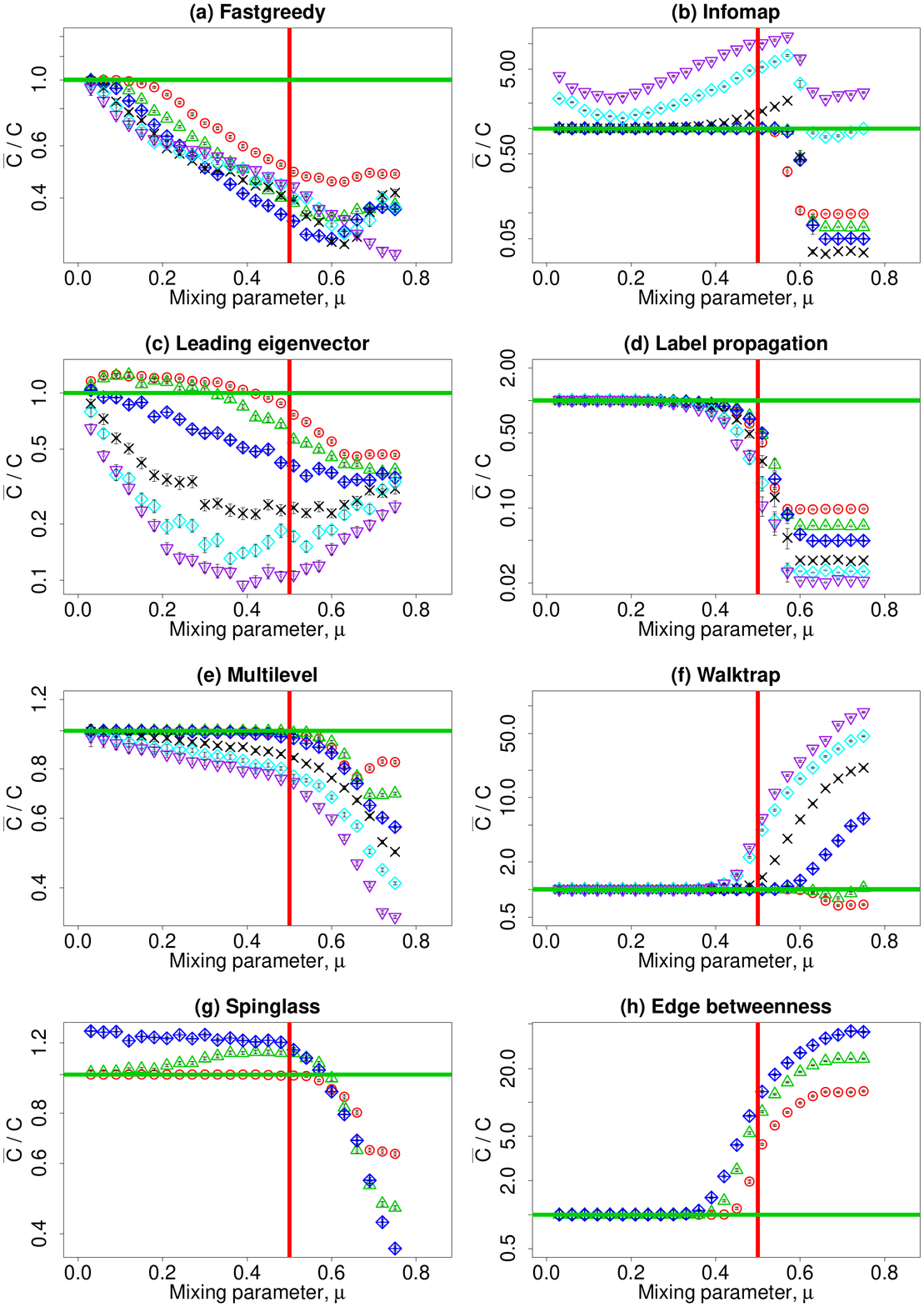}
\end{center}
\medskip
\small
Different colours refer to different number of nodes:  red ($N = 233$), green ($N = 482$), blue ($N = 1000$), black ($N = 3583$), cyan ($N = 8916$), and purple ($N = 22186$). Please notice that the vertical axis might have different scale ranges. The vertical red line corresponds to the strong definition of community where $\mu = 0.5$ and the horizontal green line represents the case that $\bar{C} = C$. The other parameters are described in Table \ref{table1}.
\label{figure2}
\end{figure}

The results for the ratio $\bar{C}/C$ as a function of the mixing parameter are shown in Figure \ref{figure2} on a \textit{log-linear} scale for all the panels. 
The Fastgreedy algorithm constantly underestimates the number of communities, and the results worsen with increasing network size and $\mu$ (Panel (a), Figure \ref{figure2}). 
For $\mu \lessapprox 0.55$, the Infomap algorithm delivers the correct number of communities of small networks ($N \lessapprox 1000$), and overestimates it for larger ones. 
For $\mu \gtrapprox 0.55$, this algorithm fails to detect any community at all for small networks and all nodes are partitioned into a single community (Panel (b), Figure \ref{figure2}). 
The leading eigenvector algorithm slightly overestimates the number of communities of small networks and the prediction worsens with increasing $\mu$. 
Moreover, it underestimates the number of communities in large networks and even the behaviour do not change monotonically with $\mu$ (Panel (c), Figure \ref{figure2}). 
The Label propagation algorithm is able to deliver the correct number of communities with small values of $\mu$ regardless of the network size. 
However, in the range $0.3 \lessapprox \mu \lessapprox 0.6$, it underestimates the number of communities and the prediction worsens with increasing network size and $\mu$. 
For $\mu \gtrapprox 0.6$, this algorithm fails to detect any community and all nodes are placed into the same community (Panel (d), Figure \ref{figure2}). 
It is apparent that the Mutilevel algorithm constantly underestimates the number of communities and such behaviour worsens with increasing network size and $\mu$ (Panel (e), Figure \ref{figure2}). 
In Figure \ref{figure2}, Panel (f), for $\mu \lessapprox 0.4$, the Walktrap algorithm delivers the correct number of communities regardless of network sizes, although the change of behaviour at which the prediction is correct depends on system size. For $\mu \gtrapprox 0.4$, this algorithm behaves differently depending on network size: it slightly underestimates the number of communities of small networks and significantly overestimates it for large ones. 
For $\mu \lessapprox 0.6$, the Spinglass algorithm constantly overestimates the number of communities, and its prediction  worsens with network size. 
When $\mu \gtrapprox 0.6$, it fails and tends to put nodes into a few giant communities (Panel (g), Figure \ref{figure2}). 
The Edge betweenness algorithm is able to deliver the correct number of communities for $\mu \lessapprox 0.4$ regardless of network size. It overestimates $C$ for $\mu \gtrapprox 0.4$ and the accuracy of the prediction worsens with increasing network size (Panel (h), Figure \ref{figure2}). 
Overall, for $\mu \lessapprox 1/2$, Infomap, Leading eigenvector, Multilevel, Spinglass, and Edge betweenness algorithms are able to deliver a reasonable estimator of the number of communities for small networks, while the number of communities obtained by Label propagation and Walktrap algorithms are relatively close to the real value regardless of network size. For $\mu \gtrapprox 1/2$, all the algorithms are much worse at detecting the correct number of communities, and among all the algorithms, Multilevel, Walktrap, and Spinglass algorithms have better outputs when the network sizes are small. 

\begin{figure}
\caption{(lower row) The mean value of the computing time of the community detection algorithms (in seconds) dependent on the mixing parameter $\mu$ on a \textit{log-linear} scale. (upper row) The standard deviation of the measures on a \textit{log-linear} scale.}
\begin{center}
\includegraphics[scale = 0.08]{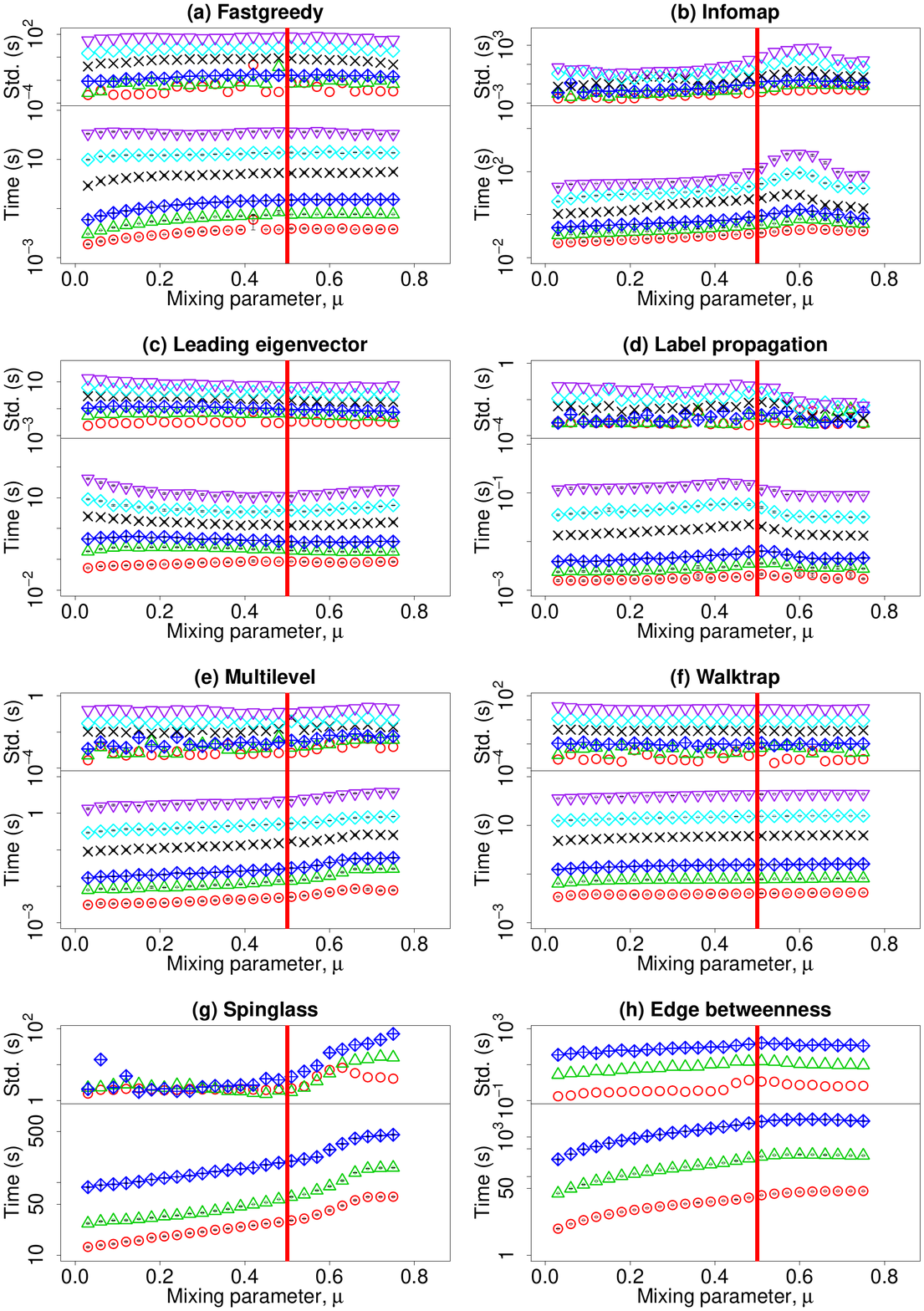}
\end{center}
\medskip
\small
Different colours refer to different number of nodes:  red ($N = 233$), green ($N = 482$), blue ($N = 1000$), black ($N = 3583$), cyan ($N = 8916$), and purple ($N = 22186$). Please notice that the vertical axis might have different scale ranges. The vertical red line corresponds to the strong definition of community where $\mu = 0.5$. The other parameters are described in Table \ref{table1}.
\label{figure3}
\end{figure}

Third, we turn to the real computing time of the algorithms. 
This measure is usually represented in theoretical estimations as a function of the number of nodes and edges. 
However, the real computing time may be also affected by the structure of the network. 
Given the number of nodes and a fixed average degree, we illustrate the computing time as a function of the mixing parameter. 
The results are shown in Figure \ref{figure3} on \textit{log-linear} scale. 
Each panel presents the computing time of a given community detection algorithm and it is subdivided in two plots: the lower one depicts the average computing time, while the upper sub-panel contains the standard deviation of the computing time when repeated over 100 different network realisations. 
Some algorithms barely depend on the mixing parameter. This is not the case for Multilevel, Spinglass, and Edge betweenness algorithms (Panel (e), (g), and (h), Figure \ref{figure3}). 
There is a slight dependency for Infomap algorithm that cannot be disregarded (Panel (b), Figure \ref{figure3}). 
The decrease of computing time for Infomap, Leading eigenvector, and Label propagation algorithms (Panel (b), (c), and (d), Figure \ref{figure3}) are accompanied with the significant worsening of NMI and $\bar{C} / C$ in Figures \ref{figure1} and \ref{figure2}. Among all the algorithms, Label propagation and Multilevel algorithms are much faster than the others (Panel (d), and (e), Figure \ref{figure3}), while Spinglass and Edge betweenness are the slowest ones (Panel (g) and (h), Figure \ref{figure3}). 

\subsection*{The observed mixing parameter}

\begin{figure}
\caption{(lower row) The mean value of the mixing parameter estimated by the community detection algorithms $\bar{\mu}$ dependent on the mixing parameter $\mu$. (upper row) The standard deviation of $\bar{\mu}$ dependent on $\mu$.}
\begin{center}
\includegraphics[scale = 0.08]{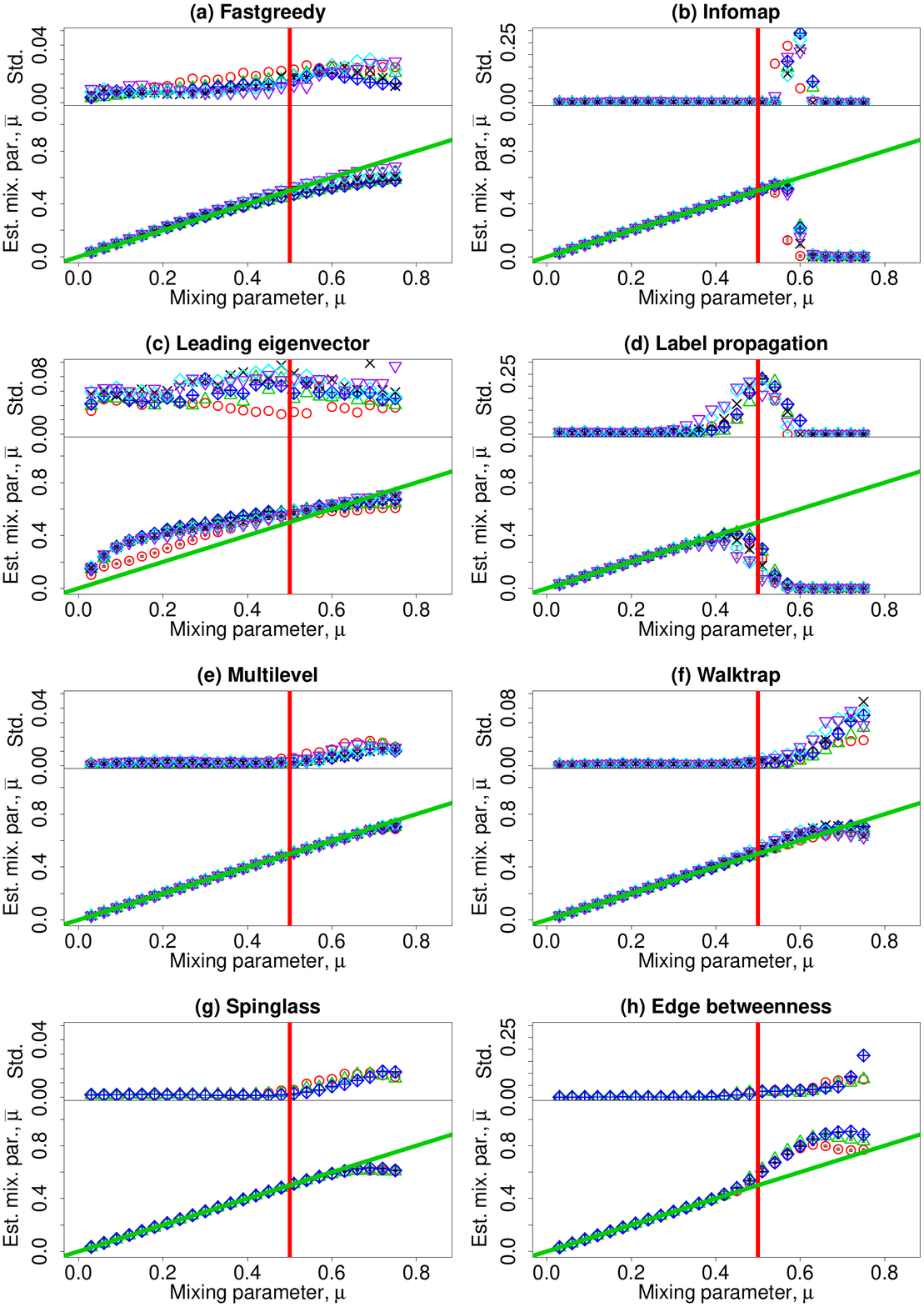}
\end{center}
\medskip
\small
Different colours refer to different number of nodes:  red ($N = 233$), green ($N = 482$), blue ($N = 1000$), black ($N = 3583$), cyan ($N = 8916$), and purple ($N = 22186$). Please notice that the vertical axis on the subfigures might have different scale ranges. The vertical red line corresponds to the strong definition of community where $\mu = 0.5$. The green line $y=x$ corresponds to the case which $\bar{\mu} = \mu$. The other parameters are described in Table \ref{table1}.
\label{figure4}
\end{figure}

Unlike the number of nodes in a network, the exact value of the mixing parameter of a graph is unobservable if ground truth is unavailable for the community assignment of nodes. 
In this section, we study the mixing parameter delivered by the community detection algorithms $\bar{\mu}$ as a function of the mixing parameter $\mu$ (see Eq. \ref{mixingparameter}). 
The results of the different algorithms are shown in the different panels of Figure \ref{figure4}. 
Each panel is subdivided in two plots: the lower has the average computed value of $\bar{\mu}$, while the upper sub-panel contains the standard deviation of the measures when repeated over 100 different network realisations. 
All algorithms have a linear (identity) relationship between $\bar{\mu}$ and $\mu$ except for the Leading eigenvector algorithm, which overshoots the results (Panel (c), Figure \ref{figure4}). 
Most of the algorithms display a turning point where the estimation of $\bar{\mu}$ breaks down. 
For the Fastgreedy, Multilevel, Walktrap, Spinglass, and Edge betweenness algorithms, $\bar{\mu}$ changes in a smooth fashion (Panel (a), (e), (f), (g), and (h), Figure \ref{figure4}). 
For the Infomap and Label propagation algorithms, the estimated mixing parameter $\bar{\mu}$ has a steep change at around $\mu \cong 0.55$ and $\mu \cong 0.5$, separately (Panel (b), and (d), Figure \ref{figure4}).   

Overall, the mixing parameter obtained by the algorithms $\bar{\mu}$ fits well with the real mixing parameter at small value of $\mu$, but it differs from the real value with increasing $\mu$. 
For certain algorithms, the estimation fails completely for larger values of $\mu$ (Infomap, Label propagation), and for the others it is either overestimated (Edge betweenness) or slightly underestimated (Fastgreedy, Walktrap, Spinglass). 
Remarkably, in the Multilevel algorithm, the estimation is very accurate for values as large as $\mu = 0.75$ for all network sizes analysed.

\subsection*{The role of network size}

\begin{figure}
\caption{(lower row) The mean value of normalised mutual information dependent on the number of nodes $N$ in the benchmark graphs on a \textit{linear-log} scale. 
(upper row) The standard deviation of the normalised mutual information dependent on $N$ on a \textit{linear-log} scale.}
\begin{center}
\includegraphics[scale = 0.08]{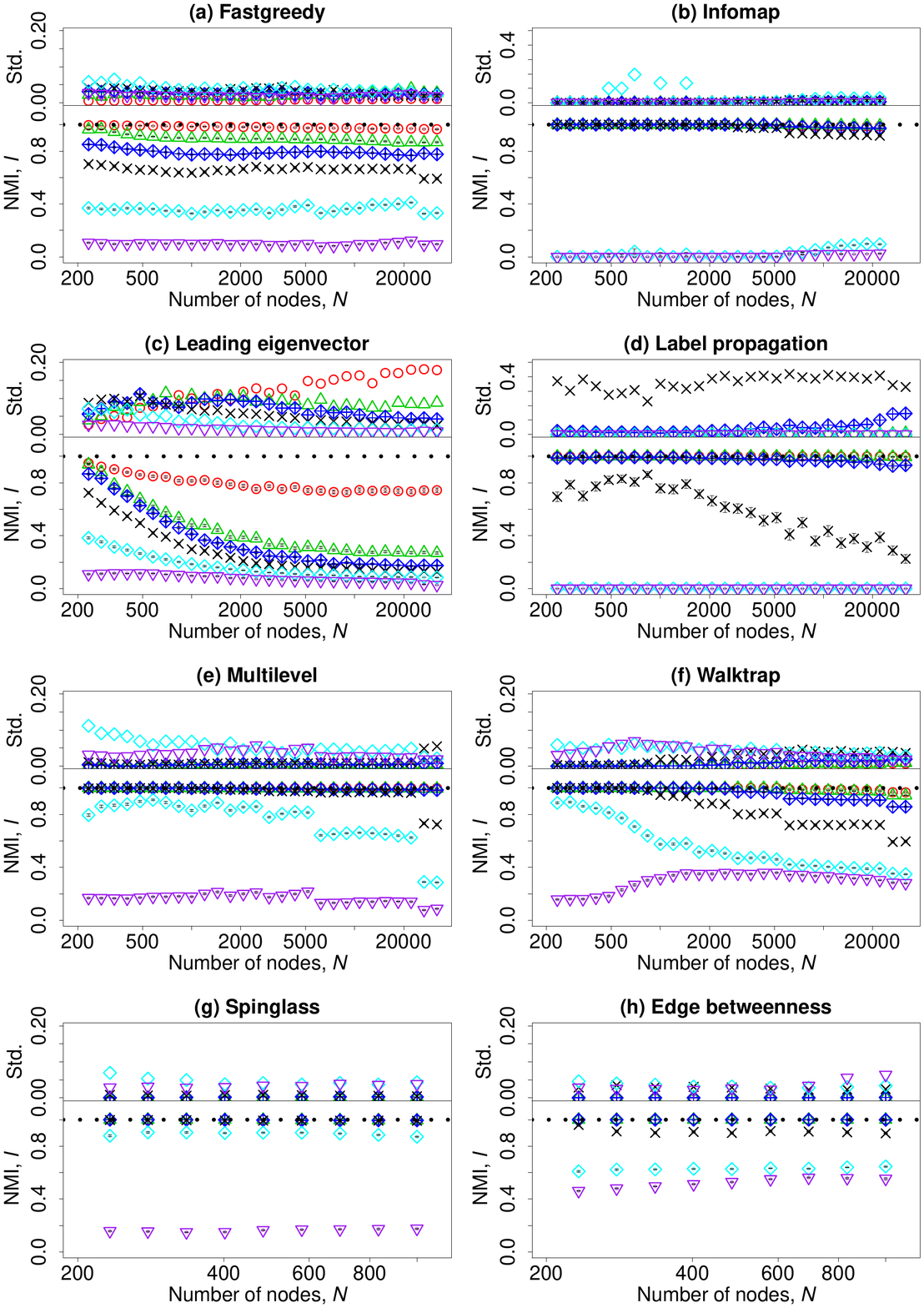}
\end{center}
\medskip
\small
Different colours refer to different values of the mixing parameter:  red ($\mu = 0.03$), green ($\mu = 0.18$), blue ($\mu = 0.33$), black ($\mu = 0.48$), cyan ($\mu = 0.63$), and purple ($\mu = 0.75$). Please notice that the vertical axis on the subfigures might have different scale ranges. The horizontal black dotted line corresponds to $I=1$. Due to the computing speed, Spinglass and Edge betweenness algorithms have been tested only on networks with $N \leq 1000$, and Infomap algorithm has been tested on networks with $N \leq 22186$. The other parameters are described in Table \ref{table1}.
\label{figure5}
\end{figure}

So far we have only discussed the role of the mixing parameter $\mu$ to the accuracy and the computing time of community detection algorithms. 
Now, as an important ingredient, we consider the effect of network size. 
In our definition of the benchmark graphs, with a fixed average degree, network size can be represented as the number of nodes in the network. 
The results are shown in Figure \ref{figure5} on a \textit{linear-log} scale. 
Each of them presents the accuracy of a given community detection algorithms and is subdivided in two plots: one for the computed value of NMI and the upped sub-panel contains the standard deviation of the measures when repeated over 100 different network realisations.
Most of the algorithms can well uncover the communities when $\mu \lessapprox 0.2$. In this case, the detecting abilities of Fastgreedy, Infomap, Label propagation, Multilevel, Walktrap, Spinglass and Edge betweenness algorithms are independent of network size (Panel (a,b,d-h), Figure \ref{figure5}). For Leading eigenvector, the accuracies decrease smoothly with network size (Panel (c), Figure \ref{figure5}). For very large $\mu \gtrapprox 0.75$, most of the algorithms fail to detect the community structure except for the Walktrap and Edge betweenness algorithms and the accuracy barely depends on network size. In the intermediate region of $\mu$, NMI is usually decreasing with network size and $\mu$.

\begin{figure}
\caption{(lower row) The mean value of the computing time of the community detection algorithms (in seconds) dependent on the number of nodes in the benchmark graphs on a \textit{log-log} scale. (upper row) The standard deviation of the computing time on a \textit{log-log} scale.}
\begin{center}
\includegraphics[scale = 0.08]{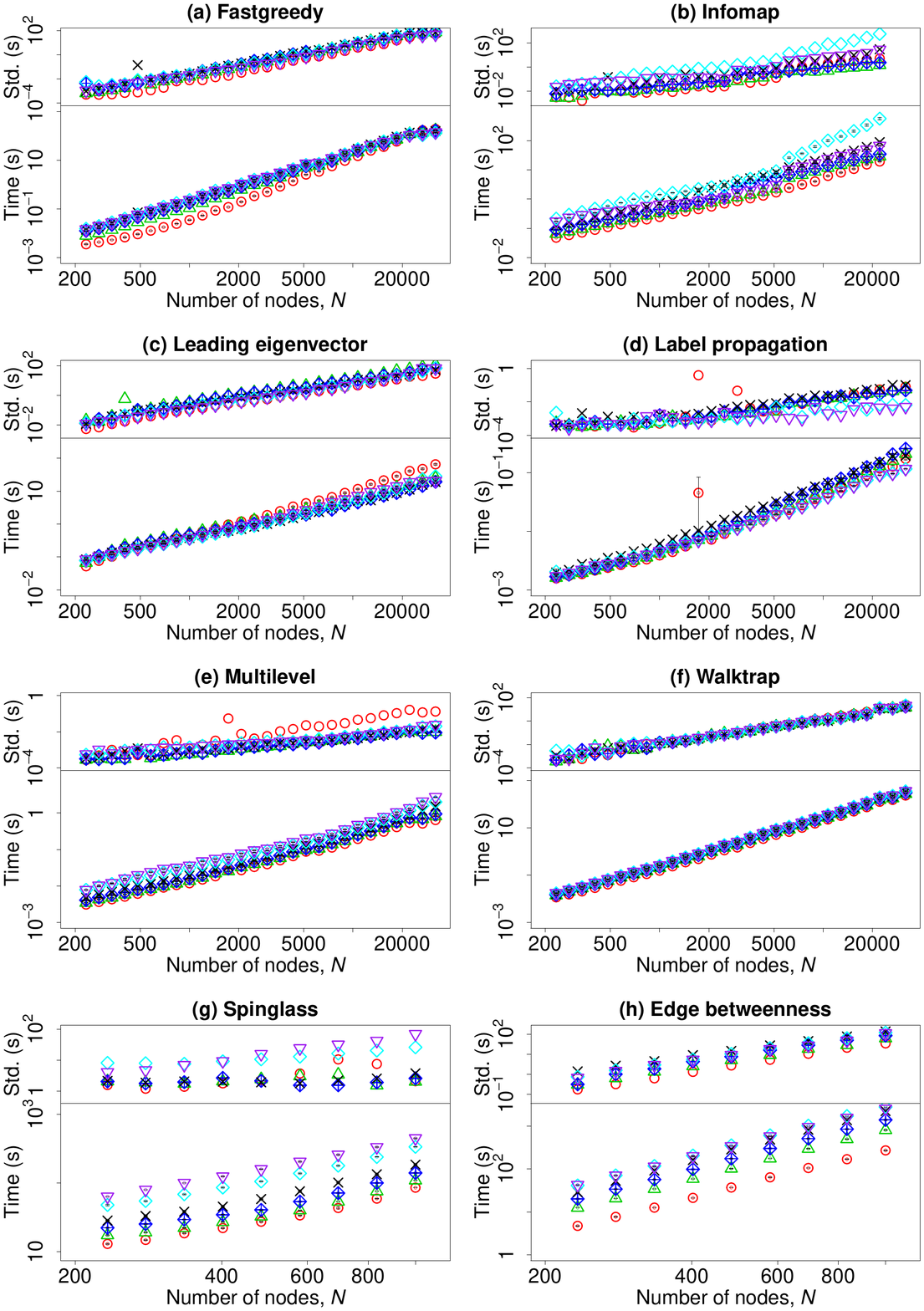}
\end{center}
\medskip
\small
Different colours refer to different values of the mixing parameter:  red ($\mu = 0.03$), green ($\mu = 0.18$), blue ($\mu = 0.33$), black ($\mu = 0.48$), cyan ($\mu = 0.63$), and purple ($\mu = 0.75$). Please notice that the vertical axis might have different scale ranges. Due to the computing speed, Spinglass and Edge betweenness algorithms have been tested only on networks with $N \leq 1000$, and Infomap algorithm has been tested on networks with $N \leq 22186$ . The other parameters are described in Table \ref{table1}.
\label{figure6}
\end{figure}

Finally, we present the computing time as a function of the network size. The results are represented in Figure \ref{figure6} on a \textit{log-log} scale. 
Each panel presents the computing time of a given community detection algorithms and is subdivided in two plots:  one for the measured value of computing time in second and the upped sub-panel contains the standard deviation of the measures when repeated over different network realisations. In the \textit{log-log} scale, there is a significant linear correlation between the computing time and the network size. To further compare the computing speed of every algorithm, we have fitted the curves according to the exponential function $T \propto N^{\alpha}$. The fitted $\alpha$ together with the corresponding adjusted R-squared values are listed in Table \ref{table2}. Only algorithms with small $\alpha$ can be applied to large networks. Overall, Label propagation algorithm is the method that scales best on network size; at the same time, Leading eigenvector, and Multilevel algorithms also have reasonable computation speeds on large networks. Fastgreedy, Infomap, Walktrap, and Spinglass algorithms scale much worse than the previous ones, and Edge betweenness algorithm is only suitable for small networks (with an almost cubic relation between network size and computing time). 

\section*{Discussion }

Traditionally, the aim of community detection in graphs has been to identify the modules by only using the information encoded in the graph topology \cite{fortunato2010community}.
In this study we have performed a comparative analysis of the accuracy and computing time of eight different community detection algorithms available in the ``igraph" package. 
Each algorithm has been tested on a set of LFR benchmark graphs 
\cite{lancichinetti2008benchmark, lancichinetti2009benchmarks}. 
The size of the benchmark graphs varies from approximately 200 to 32,000 nodes. 
With a fixed average degree, we have changed the structure of networks by using different values of the mixing parameter $\mu$. 

In this study, the limited network sizes considered here pose no challenge for modern day computers in terms of Random-Access Memory (RAM). Therefore, the memory consumption is not analysed here. However, it is worth mentioning that the maximal memory consumption could be crucial for larger scale networks: if one algorithm is implemented in a way that it needs more memory for the optimal calculation, then it can easily happen that the process slows down for large networks due to low available RAM, or it switches to a suboptimal implementation, which needs less memory. A previous study showed \cite{papadopoulos2012community} that (theoretically) many community detection methods have minimum memory consumption needs that scale linearly with the size of the graph $\mathcal{O}(2m + 2n)$, where $m$ is the number of edges and $n$ is the number of nodes. In practice, many of them need at least $\mathcal{O}(2m + 3n)$ in case of unweighted undirected graphs and when the Yale sparse matrix format is used \cite{papadopoulos2012community}. 

Our results indicate that by taking both accuracy and computing time into account, the Multilevel algorithm, which was proposed by Blondel \textit{et al.} 
\cite{blondel2008fast}, outperforms all the other algorithms on the set of benchmarks we have examined (although the modularity-based methods are known to suffer from the resolution limit of modularity \cite{fortunato2007resolution}). 
We can further apply the results in three aspects: First, since the computing time is not relevant for small networks, one should choose algorithms based their accuracies. 
Among all the algorithms, Infomap, Label propagation, Multilevel, Walktrap, Spinglass, and Edge betweenness algorithms are able to successfully uncover the structure of small networks when the mixing parameter $\mu$ is small. With increasing value of $\mu$, Infomap, Label propagation, and Edge betweenness algorithms' accuracies drop for smaller values of $\mu$ than Multilevel, Walktrap, and Spinglass algorithms. Second, for large networks, one should first choose algorithms which are able to detect the organisation of nodes in a reasonable time. In this sense, Infomap, Label propagation, Multilevel, and Walktrap algorithms are the \textit{a priori} choices. After that, by taking the accuracy into account, Multilevel is superior to the other algorithms as it displays a performance drop for a larger value of the mixing parameter $\mu$. 
Importantly, the exact value of the mixing parameter of a graph is usually unobservable. 
To get a rough idea about the value of $\mu$, one may employ either the Spinglass or the Multilevel algorithm. 
Limited by the computing time required, Spinglass algorithm cannot be applied on large networks.

\begin{figure}
\caption{Recommendation for the choice of adaptable community detection algorithms.}
\begin{center}
\includegraphics[scale = 0.15]{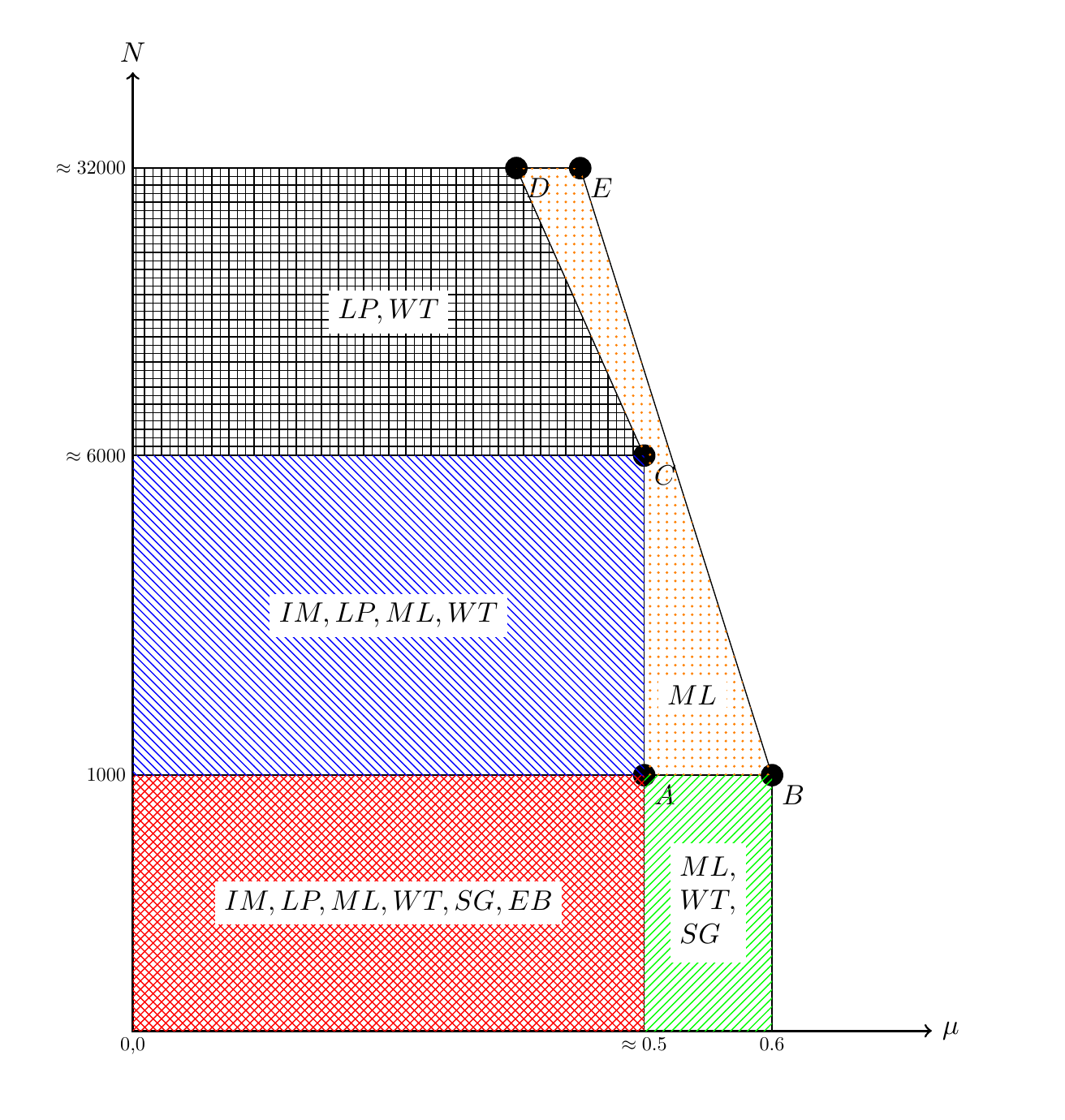}
\end{center}
\medskip
\small
The \textit{x}-axis is the mixing parameter $\mu$ and the \textit{y}-axis is the number of nodes $N$. The \textit{y}-axis is on a \textit{log} scale for better visualisation. The coordinates of certain important points are: $A (0.48, 1000)$, $B (0.6, 1000)$, $C (0.48, 6192)$, $D (0.36, 31948)$, and $E (0.42, 31948)$. In different regions we would like to recommend different algorithms, which are represented by different abbreviations: $IM$ is the Infomap algorithm, $LP$ is the Label propagation algorithm, $ML$ is the Multilevel algorithm, $WT$ is the Walktrap algorithm, $SG$ is the Spinglass algorithm, and $EB$ represents the Edge betweenness algorithm.
\label{figure7}
\end{figure}

Based on the previous results, and taking into account both factors, accuracy and computing time, it is possible to suggest under which situations to use each algorithm depending sorely on topological properties of the network under study. 
Our recommendations for the use of community detection algorithms are summarised in Figure \ref{figure7}. 
In the first region, $\mu \lessapprox 0.5$ and the network size is small, $N \lessapprox 1000$. 
There, most of the communities detection algorithms tested give accurate results (and the computing time is affordable): Infomap, Label propagation, Multilevel, Walktrap, Spinglass, and Edge betweenness can all be used in a trustworthy fashion. 
A second region has a relatively larger value of $\mu$ ($0.5 \lessapprox \mu \lessapprox 0.6$), and equally small sizes of network $N \lessapprox 1000$. There, it is possible to use Multilevel, Walktrap, and Spinglass algorithms. 
A third region encompasses again smaller values of mixing parameter ($\mu \lessapprox 0.5$) but an intermediate number of nodes ($1000 \lessapprox N \lessapprox 6000$). In this region, the best choices are Infomap, label propagation, Multilevel, and Walktrap algorithms. 
With increasing number of nodes in the networks ($6000 \lessapprox N \lessapprox 32000$), Infomap and Multilevel algorithm are very likely to provide the wrong number of communities and therefore they are no longer suitable in the fourth region. 
The last region has the highest requirement for the community detection algorithms. None of the algorithms performs very well in this region but the Multilevel algorithm outperforms all the others. 

\begin{figure}
\caption{Suggestion for the community detection process.}
\begin{center}
\includegraphics[scale = 0.12]{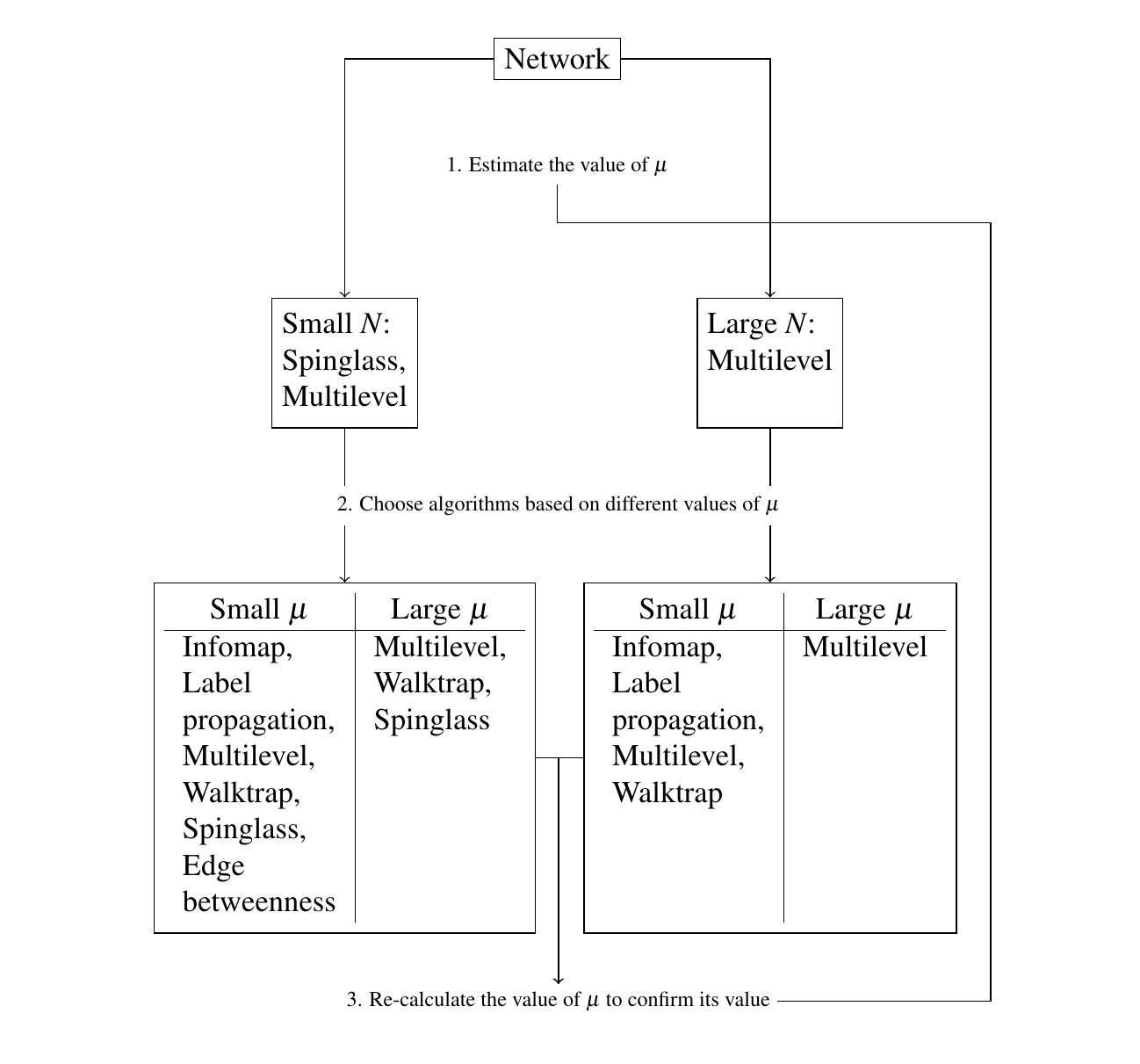}
\end{center}
\medskip
\small
Small networks are those with number of nodes less than 1000, and small $\mu$ corresponds to $\mu \lessapprox 0.5$. To be noticed that in the case that $N \ge 1000$ and $\mu \lessapprox 0.5$, Infomap and Multilevel algorithms are no longer suitable choices if $N \ge 6000$.
\label{figure8}
\end{figure}

Besides, we illustrate the suggestion for the adaptive use of the methods for community detection process in a simplified flow diagram (see Figure \ref{figure8}). With any given network, one should first employ either Spinglass algorithm or Multilevel algorithm in order to obtain an estimate of the value of the mixing parameter $\mu$. Notice that the former one can only be used for small networks ($N \lessapprox 1000$) due to the prohibitive computing time for larger network sizes. Second, one can choose a suitable method according to the values of $N$ and $\mu$ to conduct the community detection such that both the accuracy and the computing time are acceptable. Third, as we have already shown, in certain situations, there might exist large standard deviations of NMI, i.e., the community detection algorithms are not stable and therefore not reliable. Thus, the value of $\bar{\mu}$ must be recalculated to get an idea of the repeatability of the results and confirm its validity. In some situations, one might need to repeat the detection processes several times or switch to another algorithm to ensure the validity of the community detection results.   

Our suggestions have to be applied in conjunction with the concomitant research questions. As a pure application of the recommendations could bias the results. Once a researcher has decided to use a specific community detection algorithm, it is of crucial importance for her to keep in mind the limitations and the expected validity of the output of the community detection algorithm chosen.
It is noteworthy that metadata would be helpful for evaluating network community detection methods and can be used to improve the analysis and understanding of network structure \cite{yang2015defining, newman2015structure}. In real-world networks where metadata is available, researchers should also take into account the research question, the properties of the network, the interpretation and meaning of the communities while choosing the community detection algorithms. Different research questions together with the metadata might lead to different definitions of community, and further change the ground truth of the network. 

Compared to previous works on benchmarking community detection algorithms, our study has many obvious advantages: First, we have considered networks which contain a wide spectrum of number of nodes and mixing parameters. 
Second, the algorithms we have tested are integrated in a cross-platform package which has been widely used in academic research in network science and related fields. Third, we have used the LFR benchmark graphs which have shown more realistic properties than the earlier computer-generated networks such as the GN benchmark.

There are also some limitations in our work: Although the LFR benchmark has generalised the previous GN benchmark by introducing power-law distributions of degree and community size, more realistic properties are still needed. We have mainly focused on testing the effects of the mixing parameter and the number of nodes. Other properties, such as the average degree, the degree distribution exponent, and the community distribution exponent may also play a role in the comparison of algorithms. 

In the end, we stress that detecting the community structure of networks is an important issue in network science. For ``igraph" package users, we have provided a guideline on choosing the suitable community detection methods. However, based on our results, existing community detection algorithms still need to be improved to better uncover the ground truth of networks.

\section*{Methods}
In this section, we first describe in detail the procedure to obtain the benchmark networks used, then enumerate the community detection algorithms employed.

When comparing community detection algorithms, we can use either real or artificial network whose community structure is already known, which is usually termed as ground truth. 
Among the former, the celebrated Zachary's karate club 
\cite{zachary1977information} or the network of American college football teams 
\cite{girvan2002community} have been extensively used. 
Among the latter, the ones used more pervasively are the  GN 
\cite{girvan2002community} and LFR 
\cite{lancichinetti2009benchmarks} benchmarks. 
However, obtaining real networks to which a ground truth can be associated is not only difficult, but also costly in economic terms and time. 
Due to the complexity of data collection and costs, real world benchmarks usually consist of small-sized networks. 
Further, since it is not possible to control all the different features of a real network (e.g.~average degree, degree distribution, community sizes, etc.), the algorithms can only be tested -- if resorting in this kind of graphs -- on very specific cases with a limited set of features. 
In addition, the communities of real world networks are not always defined objectively or, in the best case, they rarely have a unique community decomposition. 
On the other hand, artificially generated networks can overcome most of these limitations. 
Given an arbitrary set of meso- or macroscopic properties, it is possible to generate randomly an ensemble of networks that respect them, in what is usually called generative models. 
However, as one of the most popular generative models, GN benchmark suffers from the fact that it does not show a realistic topology of the real network 
\cite{danon2006effect, lancichinetti2008benchmark} and it has very small network size. 
A recent strand of the literature on benchmark graphs tried to improve the quality of artificial networks by defining more realistic generative models: Lancichinetti \textit{et al.} extended the GN benchmark by introducing power law degree and community size distributions 
\cite{lancichinetti2008benchmark}. Bagrow had employed the Barab\'asi-Albert model 
\cite{barabasi1999emergence} rather than the configuration model \cite{molloy1995critical} to build up the benchmark graph 
\cite{bagrow2008evaluating}. Orman and Labatut proposed to use evolutionary preferential attachment model 
\cite{poncela2008complex} for more realistic properties 
\cite{orman2010effect}.

The first step to generate the LFR benchmark graph is to construct a network composed of $N$ nodes, with average degree $\hat{k}$, maximum degree $k_{max}$ and a power-law degree distribution with exponent $\alpha$ by using the configuration model.
Once this step is finished, each node has a defined total degree.
Then, given a power-law distribution of community sizes with exponent $\beta$, a set of community sizes is drawn (between arbitrarily chosen minimum and maximum values of community sizes that act as additional parameters). 
Nodes are then sequentially assigned to these communities. 
The mixing parameter $\mu$, which represents the fraction of edges a node has with nodes belonging to other communities with respect to its total degree, is the most relevant value in terms of the community structure. 
To conclude the generative algorithm, edges are rewired in order to fit the mixing parameter, while preserving the degree sequence.
This is achieved keeping fixed total degree of a node, the value of external degree is modified so that the ratio of external degree over the total degree is close to the defined mixing parameter. 
The LFR model was initially proposed to generate undirected unweighted networks with mutually exclusive communities, and was extended to generate weighted and/or directed networks, with or without overlapping communities. In this study, we focus on the undirected unweighted networks with non-overlapping communities since most of the existing community detection algorithms are designed for this type of networks. The parameter values used in our computer-generated graphs are indicated in Table \ref{table1}.\\

In this paper, we have evaluated the most widely used, state-of-the-art community detection algorithms on the LFR benchmark graphs. 
In order to make the results comparable, and reproducible, we use the implementation of these algorithms shipped with the widely used ``igraph" software package (Version 0.7.1) 
\cite{igraphinfo}. 
Here is the list of algorithms we have considered. For notation purposes when giving the computational complexity of the algorithms, the networks have $N$ nodes and $E$ edges.

\paragraph{Edge betweenness} This algorithm was introduced by Girvan \& Newman 
\cite{girvan2002community}. 
To find which edges in a network exist most frequently between other pairs of nodes, the authors generalised Freeman's betweenness centrality 
\cite{freeman1979centrality} 
to edges betweenness. 
The edges connecting communities are then expected to have high edge betweenness. 
The underlying community structure of the network will be much clear after removing edges with high edge betweenness. 
For the removal of each edge, the calculation of edge betweenness is $\mathcal{O}(E\,N)$; therefore, this algorithm's time complexity is $\mathcal{O}(E^2N)$ 
\cite{girvan2002community}.

\paragraph{Fastgreedy} This algorithm was proposed by Clauset \textit{et al.} 
\cite{clauset2004finding}. It is a greedy community analysis algorithm that optimises the modularity score. This method starts with a totally non-clustered initial assignment, where each node forms a singleton community, and then computes the expected improvement of modularity for each pair of communities, chooses a community pair that gives the maximum improvement of modularity and merges them into a new community. The above procedure is repeated until no community pairs merge leads to an increase in modularity. For sparse, hierarchical, networks the algorithm runs in $\mathcal{O}(N \, \log^2 (N))$ 
\cite{clauset2004finding}. 

\paragraph{Infomap} This algorithm was proposed by Rosvall \textit{et al.} 
\cite{rosvall2007information, rosvall2010map}. 
It figures out communities by employing random walks to analyse the information flow through a network \cite{peel2010estimating}. This algorithm starts with encoding the network into modules in a way that maximises the amount of information about the original network. Then it sends the signal to a decoder through a channel with limited capacity. The decoder tries to decode the message and to construct a set of possible candidates for the original graph. The smaller the number of candidates, the more information about the original network has been transferred. This algorithm runs in $\mathcal{O}(E)$ 
\cite{mukherjee2013dynamics}. 

\paragraph{Label propagation} This algorithm was introduced by Raghavan \textit{et al.} 
\cite{raghavan2007near}. It assumes that each node in the network is assigned to the same community as the majority of its neighbours. This algorithm starts with initialising a distinct label (community) for each node in the network. Then, the nodes in the network are listed in a random sequential order. Afterwards, through the sequence, each node takes the label of the majority of its neighbours. The above step will stop once each node has the same label as the majority of its neighbours. The computational complexity of label propagation algorithm is $\mathcal{O}(E)$ 
\cite{raghavan2007near}. 

\paragraph{Leading eigenvector} This algorithm was proposed by Newman 
\cite{newman2006finding}. The heart of this algorithm is the spectral optimisation of modularity by using the eigenvalues and eigenvectors of the modularity matrix. First, the leading eigenvector of the modularity matrix is calculated, and then the graph is split into two parts in a way that modularity improvement is maximised based on the leading eigenvector. After that, the modularity contribution is calculated at each step in the subdivision of a network. It stops once the value of the modularity contribution is not positive. Its computational complexity of each graph bipartition is $\mathcal{O}(N(E+N))$, or $\mathcal{O}(N^2 )$ on a sparse graph
\cite{xie2011community}.  

\paragraph{Multilevel} This algorithm was introduced by Blondel \textit{et al.} 
\cite{blondel2008fast}. It is a different greedy approach for optimising the modularity with respect to the Fastgreedy method. This method first assigns a different community to each node of the network, then a node is moved to the community of one of its neighbours with which it achieves the highest positive contribution to modularity. The above step is repeated for all nodes until no further improvement can be achieved. Then each community is considered as a single node on its own and the second step is repeated until there is only a single node left or when the modularity can't be increased in a single step. The computational complexity of the Multilevel algorithm is $\mathcal{O}(N \log{N})$
\cite{xie2011community}. 

\paragraph{Spinglass} This algorithm was first proposed by Reichardt \& Bornholdt 
\cite{reichardt2006statistical}. It is based on the Potts model 
\cite{wu1982potts}. The basic principle of the method is that edges should connect nodes of the same spin state (community, in the current context), whereas nodes of different states (belonging to different communities) should be disconnected. Therefore, the aim of this algorithm is to find the ground state of a spin glass model with a Potts Hamiltonian. Simulated annealing 
\cite{kirkpatrick1984optimization} has been used to minimise the system's free energy 
\cite{traag2009community}. In a sparse graph, the computational complexity of this algorithm is approximately $\mathcal{O}(N^{3.2})$ 
\cite{dahlin2013ensemble}. 

\paragraph{Walktrap} This algorithm was proposed by Pon \& Latapy 
\cite{pons2005computing}. It is a hierarchical clustering algorithm. The basic idea of this method is that short distance random walks tend to stay in the same community. 
Starting from a totally non-clustered partition, the distances between all adjacent nodes are computed. Then, two adjacent communities are chosen, they are merged into a new one and the distances between communities are updated. This step is repeated $(N-1)$ times, thus the computational complexity of this algorithm is $\mathcal{O}(E\, N^2)$. For sparse networks the computational complexity is $\mathcal{O}(N^2 \log(N))$
\cite{xie2011community}. \\

We have employed virtual machines to implement all the computation. 
For each network size and for each algorithm, a virtual machine is created using a pre-defined installation that guarantees the same execution environment conditions. The installation is tuned to guarantee that each virtual machine makes use of an entire physical node, and, at the same time, that all physical nodes where the virtual machines will be hosted have the very same hardware specifications. 
The workload distribution and collection for the results are commanded by a master-slave approach.

\section*{Acknowledgements}
The authors acknowledge financial support from the URPP Social Networks at University of Z\"urich. 
The authors are thankful to the S3IT (Service and Support for Science IT) of the University of Zurich, for providing the support and the computational resources that have contributed to the research results reported in this study, as well as Santo Fortunato for useful comments.

\section*{Author Contributions}
Z.Y., R.A. and C.J.T. designed the analysis. Z.Y. and C.J.T. devised the methodology. Z.Y. analysed the data. Z.Y., R.A. and C.J.T. wrote the manuscript.

\section*{Competing interests}
The authors declare no competing financial interests.

\begin{table}[h]
\caption{Parameters of LFR benchmark graphs.}
\begin{center}
\begin{tabular}{ l | l }
\hline \hline
Parameter & Value \\ 
\hline
Number of nodes $N$ & 233 $\sim$ 31948 \\  
\hline
Maximum degree & $0.1 N$\\
\hline
Maximum community size & $0.1 N$\\
\hline
Average degree & 20\\
\hline
Degree distribution exponent & -2\\
\hline
Community size distribution & -1\\
exponent & \\
\hline
Mixing coefficient $\mu$ & [0.03, 0.75]\\
\hline
\end {tabular}
\end{center}
\medskip
\small
To deal with possible discrepancies in the network properties, we have randomly generated 100 network for every set of parameters. Due to the slow computing speed, Spinglass and Edge betweenness algorithms have been tested only on small networks with $N \leq 1000$.
\label{table1}
\end{table}

\begin{table}
\caption{
Indexes of the exponential function $T  \propto N^{\alpha}$ with the corresponding adjusted R-squared values.}
\begin{center}
\begin{tabular}{ c | c | c | c | c}
\hline \hline
 & Fastgreedy & Infomap & \begin{tabular}{@{}c@{}}Leading \\ eigenvector\end{tabular} 
& \begin{tabular}{@{}c@{}}Label \\ propagation\end{tabular}\\ 
\hline
$\alpha$ & 2.048 [0.006] & 1.421 [0.009] & 1.123 [0.005] & 0.959 [0.005]\\
\hline
$R^2$ & 0.956 & 0.933 & 0.951 & 0.947\\
\hline
 & Multilevel & Walktrap & Spinglass & 
\begin{tabular}{@{}c@{}}Edge \\ betweenness\end{tabular}\\ 
\hline
$\alpha$ & 1.126 [0.003] & 2.04 [0.002] & 1.282 [0.013] () & 2.915 [0.005]\\
\hline
$R^2$ & 0.957 & 0.962 & 0.867 & 0.884\\
\hline
\end {tabular}
\end{center}
\medskip
\small
The standard errors are listed in brackets. All the results are statistically significant at the significance level of 0.05. Spinglass and Edge betweenness algorithms have been tested only on small networks with $N \leq 1000$, there might be some biases in the indexes of these two methods.
\label{table2}
\end{table}

\end{document}